\documentclass{aa}  
\usepackage{natbib}
\usepackage{graphicx}
\usepackage{txfonts}
\usepackage{hyperref}
\usepackage{xcolor}

\newcommand{\planck}{\textsl{Planck}}
\newcommand{\rosat}{\textsl{ROSAT}}
\newcommand{\xmm}{\textsl{XMM-Newton}}
\newcommand{\suzaku}{\textsl{Suzaku}}
\newcommand{\beq}{\begin{equation}}
\newcommand{\eeq}{\end{equation}}
\newcommand{\beqa}{\begin{eqnarray}}
\newcommand{\eeqa}{\end{eqnarray}}
\newcommand{\mpc}{$h^{-1} \mathrm{Mpc}$}
\def\msun{{\rm M}_{\odot}}

\begin{document} 

  \title{Detection of intercluster gas in superclusters \\ using the thermal Sunyaev-Zel'dovich effect}
  \author{H. Tanimura\inst{1} \and N. Aghanim\inst{1} \and M. Douspis\inst{1} \and A. Beelen\inst{1} \and V. Bonjean\inst{1,2}}

  \institute{
  Institut d’Astrophysique Spatiale, CNRS (UMR 8617), Université Paris-Sud, Bâtiment 121, Orsay, France \and
  LERMA, Observatoire de Paris, PSL Research University, CNRS, Sorbonne Universités, UPMC Univ. Paris 06, 75014, Paris,
France \\
              \email{hideki.tanimura@ias.u-psud.fr}
             }
             
  \date{}

  \abstract 
        {Using a thermal Sunyaev-Zel'dovich (tSZ) signal, we search for hot gas in superclusters identified using the Sloan Digital Sky Survey Data Release 7 (SDSS/DR7) galaxies. We stack a Comptonization $y$ map produced by the \planck\ Collaboration around the superclusters and detect the tSZ signal at a significance of 6.4$\sigma$. We further search for an intercluster component of gas in the superclusters. For this, we remove the intracluster gas in the superclusters by masking all galaxy groups/clusters detected by the \planck\ tSZ, \rosat\ X-ray, and SDSS optical surveys down to a total mass of $10^{13} \, \msun$.  We report the first detection of intercluster gas in superclusters with $y = (3.5 \pm 1.4) \times 10^{-8}$ at a significance of 2.5$\sigma$. Assuming a simple isothermal and flat density distribution of intercluster gas over superclusters, the estimated baryon density is $(\Omega_{\rm gas} / \Omega_{\rm b}) \times (T_{\rm e}/8 \times 10^{6} \, {\rm K}) = 0.067 \, \pm 0.006 \, \pm 0.025$. This quantity is inversely proportional to the temperature, therefore taking values from simulations and observations, we find that the gas density in superclusters may account for 17 - 52\% of missing baryons at low redshifts. A better understanding of the physical state of gas in the superclusters is required to accurately estimate the contribution of our measurements to missing baryons. }
        

\maketitle

%_______________________________________________________

\section{Introduction}
%_______________________________________________________

The formation of cosmic web structure composed of voids, filaments, and clusters of galaxies is expected in the standard cosmological model of structure formation (e.g., \citealt{Zeldovich1982}). Superclusters of galaxies are the largest over-dense regions in the Universe extending over tens of megaparsecs (e.g., \citealt{Reisenegger2000, batiste2013, mill2015}). Superclusters may become bound isolated structures or rather may not be gravitationally bound and split into several systems in the future \citep{araya2009, chon2015}. Their formation and evolution, similar to those of the cosmic web, are governed by dark matter and dark energy and superclusters can thus be used to test cosmological models. 

Galaxies and clusters of galaxies are concentrated in superclusters and they can be identified by multiple systems of galaxy clusters or density enhancements of galaxy distribution. For example, superclusters are identified on the basis of Abell clusters \citep{einasto2001} and galaxy groups \citep{einasto2007} using smoothed density field. The friends-of-friends (FoF) method can also be used to identify superclusters from Abell clusters \citep{chow2014} and from SDSS galaxy samples \citep{basilakos2003}. However, the number of known superclusters is still small at present, especially for the number of very large superclusters. 

Superclusters have complex inner structures and are excellent laboratories to study the properties and evolution of galaxies and clusters of galaxies. \cite{proust2006} suggests that intercluster galaxies in the Shapley Supercluster might contribute up to twice as much mass to the supercluster as cluster galaxies. Furthermore, \cite{einasto2011} shows that high-density cores of galaxy clusters are connected by galaxy chains of filaments in superclusters. However, these are based on the studies with a stellar component comprising $\lesssim$ 10\% baryons. Most of baryons exist as gas components, for which the distribution is unknown.

The so-called warm hot intergalactic medium (WHIM), with a temperature range of $10^5$--$10^7$ K \citep{Cen2006}, is difficult to observe due to its low density. At high redshift ($z \gtrsim 2$), most of the expected baryons are found in the Ly$\alpha$ absorption forest: the diffuse, photo-ionized intergalactic medium (IGM) with a temperature of $10^4$--$10^5$ K (e.g., \citealt{Weinberg1997, Rauch1997}).   However, at redshifts $z \lesssim 2$, the observed baryons in stars, the cold interstellar medium, residual Ly$\alpha$ forest gas, OVI and BLA absorbers, and hot gas in clusters of galaxies account for only $\sim$50\% of the expected baryons -- the remainder has yet to be identified (e.g., \citealt{Fukugita2004, Nicastro2008, Shull2012}). Hydrodynamical simulations suggest that 40--50\% of baryons could be in the form of shock-heated gas in the cosmic web between clusters of galaxies. Several detections in far-UV and X-ray have been reported, but few are considered definitive \citep{Yao2012}.

Large amounts of missing baryons may be encompassed in a crowded environment of superclusters.  The identification of inner structures can be an effective way to search for the elusive WHIM. A search for filamentary connections between clusters in the Shapley supercluster was performed by \cite{kull1999} using the \rosat\ data. They detected a diffuse X-ray emission in 0.5-1 keV between Abell 3558 and Abell 3556. However, the projected position places the region within virial radius of both clusters and makes it difficult to claim that the signal is associated with a nonvirialized filament of moderate density.  Moreover, \cite{rines2001} demonstrated that Abell 2199 supercluster is kinematically connected to Abell 2197 and one or two X-ray emitting systems, and also identified five X-ray faint groups between them. This may suggest the existence of intercluster gas in the extended filament between them. 

The thermal Sunyaev-Zel'dovich (tSZ) effect \citep{Zeldovich1969, Sunyaev1970, Sunyaev1972, Sunyaev1980} arises from the Compton scattering of CMB photons as they pass through hot ionized gas along the line of sight.  The signal provides an excellent tool for probing baryonic gas at low and intermediate redshifts. Electron pressure in the WHIM would be sufficient to generate potentially observable tSZ signals.  However, the measurement is challenging due to the morphology of the source and the relative weakness of the signal. Some detections of the tSZ signal from filamentary structures are reported in \cite{Planck2013IR-VIII} and \cite{bonjean2018} and statistically by stacking methods in \cite{deGraaff2017} and \cite{Tanimura2019}, but the results may be affected by the fact that properties of filaments (shape, density, temperature, etc.) are not well understood. 

The \planck\ collaboration has produced a full-sky tSZ map (Comptonization $y$ map) with 10 arcmin angular resolution and high sensitivity \citep{Planck2014-XXI, Planck2016-XXII}.  In addition to numerous galaxy clusters detected in the \planck\ data \citep{planck2011er-viii, planck2014-xx, planck2016-XXIV}, the \planck\ collaboration reports the first significant tSZ signal from superclusters \citep{planck2011er-ix}. With a followup study of \xmm , PLCK G214.6+37.0 is found to be the most massive and X-ray brightest with triple systems of galaxy clusters. A cross-correlation with SDSS-DR7 luminous red galaxies (LRG) and SDSS-DR7 superclusters \citep{Liivamagi2012} suggested that this triple system is encompassed in a very-large-scale structure located at z $\sim$ 0.45 \citep{planck2013ir-vi}, as part of supercluster structure. These multi-frequency studies shows that only $\sim$68 \% of the total tSZ signal can be explained by the predictions from the X-ray signal. The discrepancy may hint at the presence of diffuse intercluster gas in the supercluster. 

In this paper, we probe a hot gas in superclusters through the tSZ effect. The following datasets are used in our analysis: SDSS DR7 supercluster catalog, \planck\ $y$ map, and several galaxy cluster catalogs are used to construct a mask in Section 2. In Section 3, we present a stacking method employed since the signal-to-noise ratio in the \planck\ $y$ map is not high enough to trace superclusters individually. In Section 4, possible systematic effects and interpretations of our measurements are discussed. Finally, we summarize our findings in Section 5. Throughout this work, we adopt a $\Lambda$CDM cosmology from Table 4 (TT, TE, EE + lowP + lensing + ext) in \cite{Planck2016-xiii}. Masses are quoted in solar mass and $M_{\Delta}$ is the mass enclosed within a sphere of radius $R_{\Delta}$ such that the enclosed density is $\Delta$ times the critical density at redshift $z$.

%_______________________________________________________

\section{Data}
%_______________________________________________________

\subsection{Planck $y$ maps}
The \planck\ $y$ map is one of the datasets provided in the \planck\ 2015 data release.  It is available in HEALpix\footnote{http://healpix.sourceforge.net/} format \citep{gorski2005} with a pixel resolution of $N_{\rm side}$ = 2048.  Two types of $y$ map are publicly available: MILCA \citep{hurier2013} and NILC \citep{remazeilles2013}, both of which are based on multi-band combinations of the \planck\ frequency maps \citep{Planck2016-XXII}. The $y$ map produced with NILC shows a higher noise level at large scales \citep{Planck2016-XXII}. Such large-scale noise can be difficult to model precisely in a stacking analysis for superclusters subtending relatively large angular scales. For this reason, we base our analysis on the \planck\ $y$ map produced with MILCA and we check the consistency of our results with the NILC $y$ map in Sect. \ref{sec:interp}.

The 2015 \planck\ data release also provides sky masks suitable for analyzing the $y$ maps, including a point-source mask and galactic masks that exclude 40, 50, 60, and 70\% of the sky. We combine the point source mask with the 40\% galactic mask which excludes $\sim$50\% of the sky (upper panel in Fig.~\ref{f3}).

\subsection{Supercluster catalog}
\label{subsec:sc-catalog}
The SDSS DR7 supercluster catalog is constructed from flux-limited samples of the 583 362  SDSS DR7 spectroscopic galaxies at $z<0.2$ \citep{Liivamagi2012}. The superclusters are defined as over-dense regions in the smoothed luminosity density field using the $B_{3}$ spline kernel with a radius of 8 \mpc . Two types of threshold are used, one with an adaptive local threshold and the other with a global threshold. We adopt the ``main'' catalog of  982 well-defined superclusters with a high global threshold to study the details of the structure. The catalog provides two central positions: one using the peak in the smoothed luminosity density field (Luminosity peak) and the other deduced from the centroid of the luminosity field (Luminosity center). The catalog also provides their diameters defined by the maximum distance between its member galaxies as well as distances to the central positions from us. The supercluster volumes are also estimated by summing grid cells in the luminosity density grid above a threshold.

\subsection{Catalog of galaxy groups/clusters}
\label{subsec:groups}

\begin{figure}
    \centering
    \includegraphics[width=\linewidth]{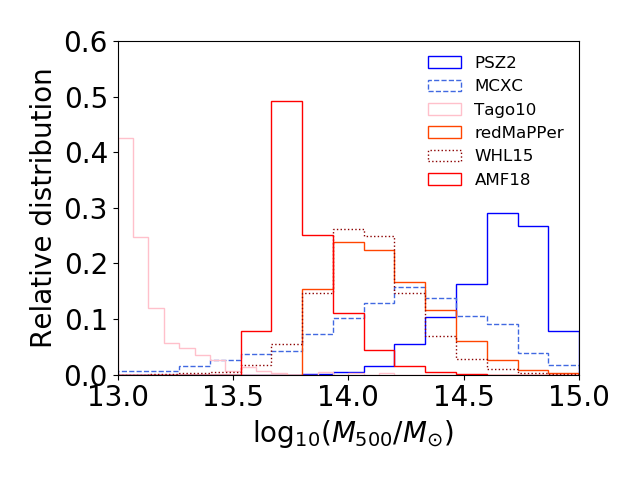}
    \includegraphics[width=\linewidth]{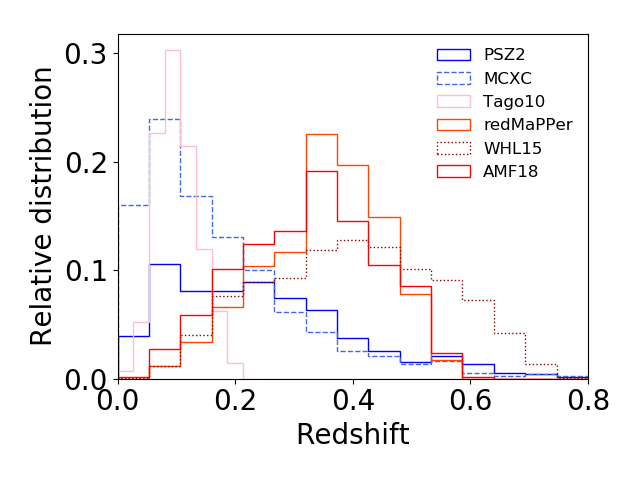}
    \caption{{\it Upper}: Mass distribution of galaxy groups/clusters masked in our analysis. The catalogs of these groups/clusters are described in Sect. \ref{subsec:groups}.  {\it Lower:} Redshift distribution of the galaxy groups/clusters. We note that the catalog from Tago10 is constructed in $z<0.2,$ and therefore the number of bins for Tago10 is doubled to see others clearly. } 
    \label{f1}
    \end{figure}

We briefly present galaxy group/cluster catalogs used in our analysis. In order to detect the contribution from intercluster gas in superclusters to the tSZ signal, all clusters and groups from the catalogs listed below are masked in the \planck\ $y$ map. In Fig.~\ref{f1}, we show the mass and redshift distribution of all the galaxy groups/clusters used in our analysis.

The \planck\ collaboration constructed a catalog of galaxy clusters detected using the tSZ effect from the 29 months of full-mission data \citep{Planck2016-XXVII} (PSZ2). This contains 1653 sources, of which 1203 are confirmed clusters from a multi-wavelength search for counterparts with radio, microwave, infrared, optical, and X-ray data sets. The masses ($M_{500}$) of 1094 clusters with redshifts are estimated with the SZ flux using the scaling relation $Y_{500} - M_{500}$ from \cite{planck2014-xx}. 

The MCXC is the catalog of galaxy clusters based on the \rosat\ All Sky Survey \citep{Piffaretti2011} (MCXC). The MCXC comprises 1743 clusters, for which total mass ($M_{500}$) is estimated using the $L_{500} - M_{500}$ scaling relation from \cite{Pratt2009} as well as the radius ($R_{500}$). 

\cite{Tago2010} extracts 78,800 groups of galaxies from the same SDSS DR7 galaxies used in \cite{Liivamagi2012}, by adopting a modified friends-of-friends method with a slightly variable linking length. The virial radii are estimated from Eq. (3) in \cite{Tago2010} using projected distances between member galaxies. 

\cite{Rykoff2014} introduces redMaPPer, which is a red-sequence cluster finder designed to make optimal use of large photometric surveys (redMaPPer). They apply the redMaPPer algorithm to $\sim$ 10,000 deg$^2$ of SDSS DR8 data and present the resulting catalog of 26,111 clusters (redMaPPer catalog v6.3) over the redshift range between 0.08 and 0.55. The algorithm exhibits excellent photometric redshift performance and the richness estimates are tightly correlated with external mass proxies. We estimate the masses ($M_{500}$) using the scaling relation of $M_{500} - \lambda$ described in \cite{Jimeno2017}. 

\cite{Wen2012} identify 132,684 groups and clusters using photometric redshifts of galaxies from the SDSS DR8 data in the redshift range between 0.05 and 0.8 (WHL12). The catalog has been updated with 25,419 new rich clusters at high redshift using the SDSS DR12 spectroscopic data in \cite{Wen2015} (WHL15). To determine the masses of galaxy clusters accurately, the masses of 1191 clusters estimated by X-ray or tSZ measurements are used to calibrate the optical mass proxy, in which they find that the masses ($M_{500}$) are well correlated with the total luminosity ($L_{500}$). Using the scaling relation, the masses of all 158,103 clusters have been updated, of which 89\% have spectroscopic redshifts. 

\cite{Banerjee2018} present a galaxy cluster catalog constructed from the SDSS DR9 data using the Adaptive Matched Filter (AMF) technique \citep{Kepner2000} (AMF18). The catalog has 46,479 galaxy clusters with richness $\Lambda_{200} > 20$ in the redshift range from 0.045 to 0.641 in $\sim$ 11,500 deg$^2$ of the sky. The AMF algorithm identifies clusters by finding peaks in a cluster likelihood map generated from galaxy positions, magnitudes and redshifts. The NFW density profile is assumed in AMF18 to construct the cluster likelihood map. The AMF approach provides a simultaneous determination of richness, core and virial radii ($R_{200}$), and redshift. 

In \cite{Banerjee2018}, the AMF18 catalog is compared with the redMaPPer (26,350 clusters) and WHL12 (132,684 clusters) catalogs in the same area of the sky and in the overlapping redshift range. The AMF18 clusters match 97\% of the richest Abell clusters (Richness group 3), as in WHL15, while the redMaPPer clusters match $\sim$ 90\% of those clusters. For comparisons with X-ray clusters such as MCXC and REFLEX, especially for most luminous clusters ($Lx > 8 \times 10^{44} \, \mathrm{ergs/sec}$),  \cite{Banerjee2018} finds that AMF18 performs equivalently to WHL15 for identifications of clusters.

    \begin{figure}
    \centering
    \includegraphics[width=\linewidth]{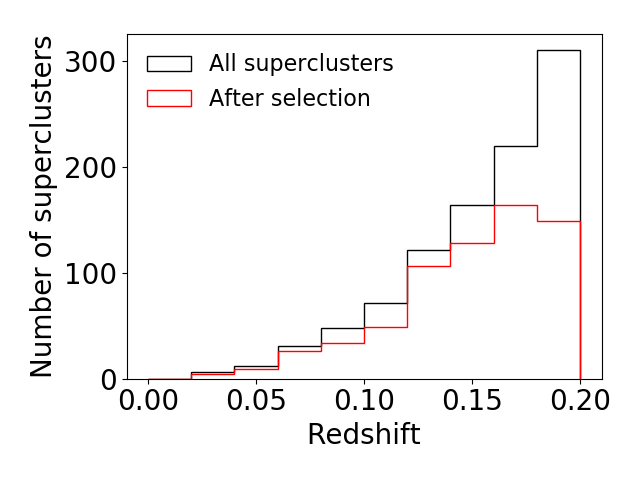}
    \includegraphics[width=\linewidth]{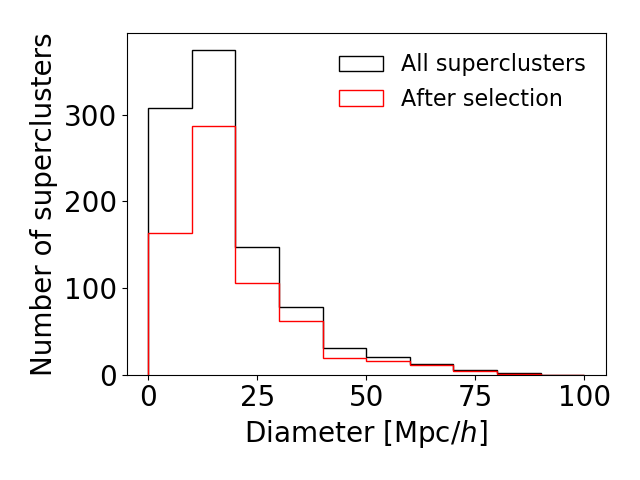}
    \caption{{\it Upper}: Redshift distribution of superclusters.  {\it Lower:} Diameter distribution of superclusters in Mpc/$h$. The {\it black} and {\it red} distributions contain either all the 982 superclusters in the catalog or the 669 selected superclusters used in our stacking analysis, respectively.}
    \label{f2}
    \end{figure}
    
    \begin{figure}
    \centering
    \includegraphics[width=\linewidth]{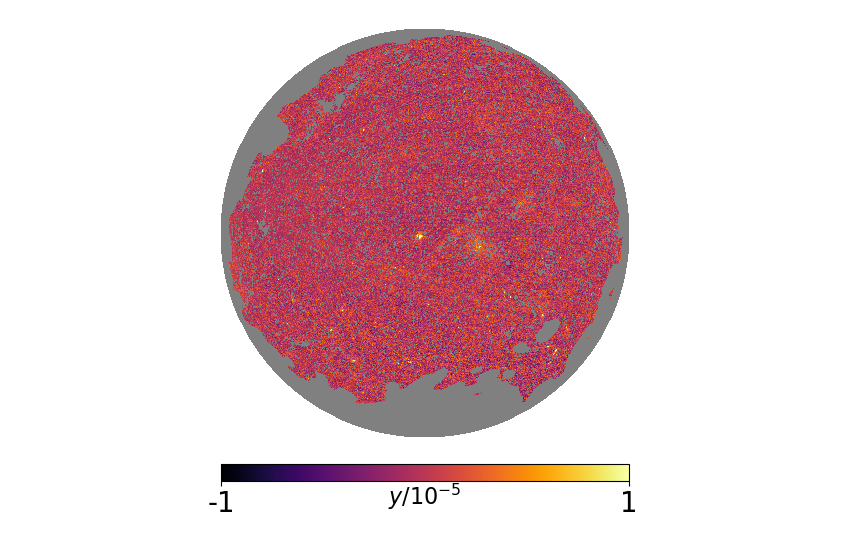}
    \includegraphics[width=\linewidth]{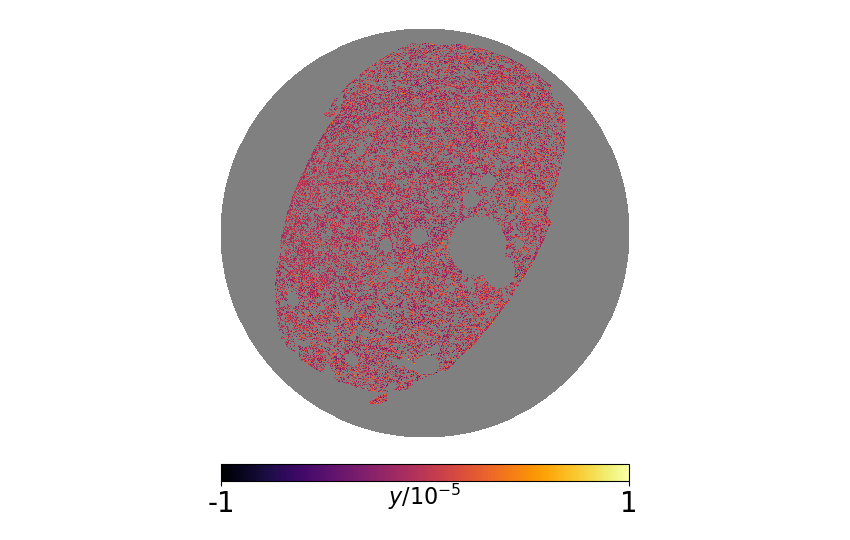}
    \caption{{\it Upper}: \planck\ $y$ map with the 40\% galactic mask and point-source mask from the view of the north galactic pole. {\it Lower:} \planck\ $y$ map after masking galaxy clusters. The galaxy groups/clusters detected by the \planck\ tSZ, \rosat\ X-ray, and SDSS optical surveys described in Sect. \ref{subsec:groups} are all masked by three times the radius ($3 \times R_{500}$) of each galaxy cluster. The region outside of the SDSS DR7 survey is also excluded. }
    \label{f3}
    \end{figure}

%_______________________________________________________

\section{Analysis}
%_______________________________________________________

\subsection{Stacking $y$ map centered on superclusters}
\label{subsec:stacking}
In this section, we describe our procedure for stacking the \planck\ $y$ map at the positions of the superclusters and construct the stacked $y$ profile. For each supercluster, we extract the $y$ map on the same grid in ``scaled radius'' in a 2D coordinate system of $-2.5 < \Delta l/\theta_{sc} < +2.5$ and $-2.5 < \Delta_b/\theta_{sc} < +2.5$ divided into 31 $\times$ 31 bins and the corresponding $y$ profile, where $\theta_{sc}$ is the angular radius of the superclusters. The scaled radius is calculated for each supercluster using the half diameter and radial distance provided in the catalog. The projected distances on the \planck\ $y$ map are normalized accordingly. The mean tSZ signal in an annular region of [1.5, 5.0] $\times$ (an angular size of supercluster) is subtracted for each supercluster as the local background signal. 

\begin{figure*}
    \centering
    \includegraphics[width=0.49\linewidth]{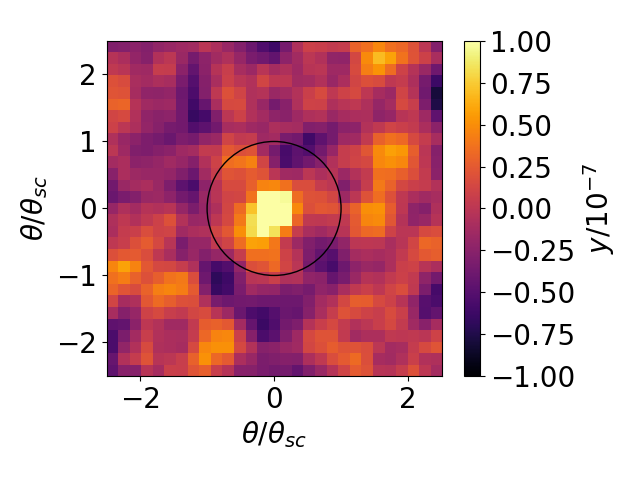}
    \includegraphics[width=0.49\linewidth]{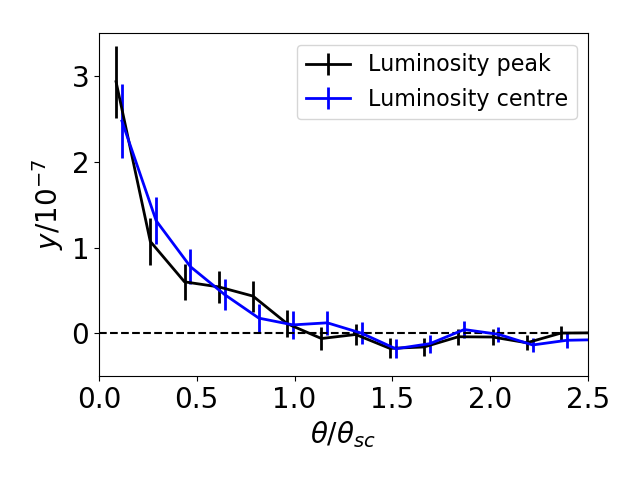}
    \caption{{\it Left}: Average \planck\ $y$ map stacked against 790 superclusters in a coordinate system where the superclusters are located at the center and the sizes are normalized to one. For centeres of superclusters, ``luminosity peak'' is used. The square region, $-2.5 < \Delta_l/\theta_{sc} < +2.5$ and $-2.5 < \Delta_b/\theta_{sc} < +2.5$, comprises 31 $\times$ 31 pixels. The {\it black} circle represents a boundary of superclusters assuming circular shapes. {\it Right}: Corresponding radial $y$ profile ({\it black}) is compared with the $y$ profile using ``luminosity center'' ({\it blue}). The 1$\sigma$ uncertainties are estimated by a bootstrap resampling (see Sect. \ref{subsec:uncertainties}).}
    \label{f4}
    \end{figure*}

First, we stack the $y$ map without masking the galaxy clusters (upper panel in Fig.~\ref{f3}). In this step, we analyze the stacked signal of a sample of 790 superclusters from the catalog of \cite{Liivamagi2012}, laying outside the \planck\ galactic and point-source masks. Superclusters laying at the boundary of the SDSS survey are also discarded since the central positions may not be clearly determined. The left panel in Fig.~\ref{f4} shows the average stacked  ``background-subtracted'' $y$ map and the right panel shows the corresponding $y$ profile with 1$\sigma$ uncertainties. The uncertainties are estimated by a bootstrap resampling (see Sect. \ref{subsec:uncertainties}). The tSZ signal is detected at a significance of 6.4$\sigma$ and is dominated by the central peak of $y \sim 2.9 \times 10^{-7}$. This shows that hot gas (mainly in galaxy clusters) traced by the tSZ signal is concentrated in the central positions of the superclusters. 

Two centers, luminosity peak and luminosity center, are defined in the catalog of superclusters. The stacked $y$ profiles using these two centers are shown in the right panel of Fig.~\ref{f4}. These $y$ profiles are consistent with each other and the ``luminosity peak'' as well as the ``luminosity center'' correspond to the peak of the tSZ signal. This indicates that both of these centers should trace the center of the associated gravitational potential. In our analysis, we choose luminosity peak as a supercluster center; it gives the highest tSZ amplitude. We find that consistent results are obtained using luminosity center as a supercluster center.

\subsection{Applying the mask of galaxy clusters}
\label{subsec:masksize}
Now, we focus on probing intercluster gas, located outside of galaxy clusters in superclusters. We therefore mask the galaxy groups and clusters identified in the \planck\ tSZ, \rosat\ X-ray, and SDSS optical surveys described in Sect. \ref{subsec:groups}. We apply the cluster mask with three times the cluster radius ($3 \times R_{500}$) in size. For \planck\ SZ clusters without assigned radius, we mask a region of 10 arcmin in radius, corresponding to the beam size of the \planck\ $y$ map. In addition, we remove the region outside of the SDSS DR7 survey. The lower panel in Fig.~\ref{f3} shows the $y$ map we use to probe the intercluster gas. Eventually, the effective area for our analysis is 7.7\% of the sky.

\begin{figure}
    \centering
    \includegraphics[width=\linewidth]{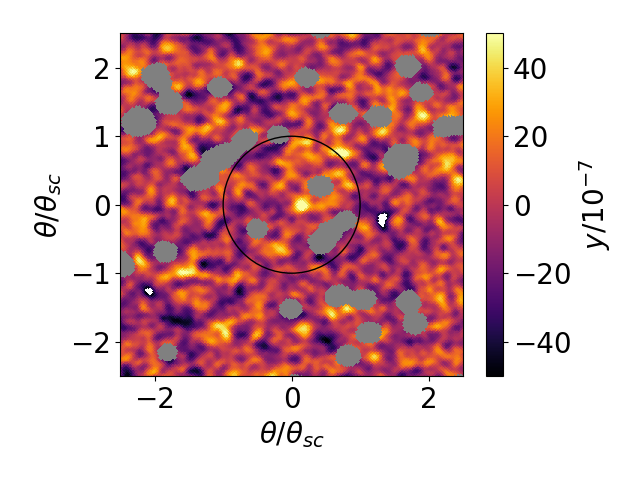}
    \includegraphics[width=\linewidth]{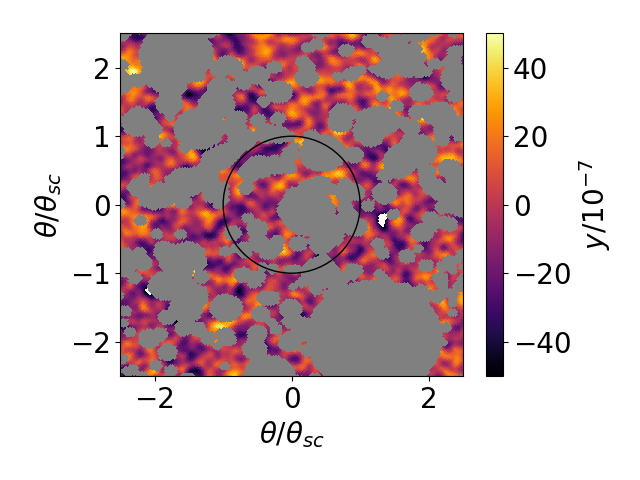}
    \caption{The $y$ map around one supercluster before masking galaxy clusters ({\it upper}) and after masking them ({\it lower}). The mask size is set to be three times the radius ($3 \times R_{500}$) of galaxy clusters. }
    \label{f5}
    \end{figure}

The union mask is applied during the stacking process: for a given supercluster, the masked pixels in the \planck\ $y$ map are not accumulated in the stacked image. As an example, one supercluster is shown before masking galaxy clusters and after masking them in Fig.~\ref{f5}. Without the mask, bright signals from galaxy clusters were seen especially around the core, but they are well covered by the mask. 

Due to this mask, some superclusters are largely masked and may bias our results. Therefore, we discard superclusters  from our analysis if 1) the available region is less than 20\% or 2) less than $\sim$ 0.3 $\times$ 0.3 deg$^2$. Here again, the superclusters laying at the boundary of the SDSS survey are discarded from our analysis. We finally perform the stacking on the remaining  669 superclusters. We find consistent results using different selection criteria.

\begin{figure*}
    \centering
    \includegraphics[width=0.33\linewidth]{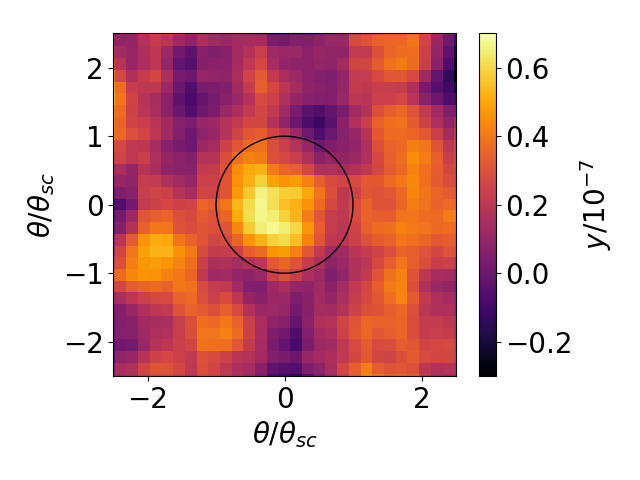}
    \includegraphics[width=0.33\linewidth]{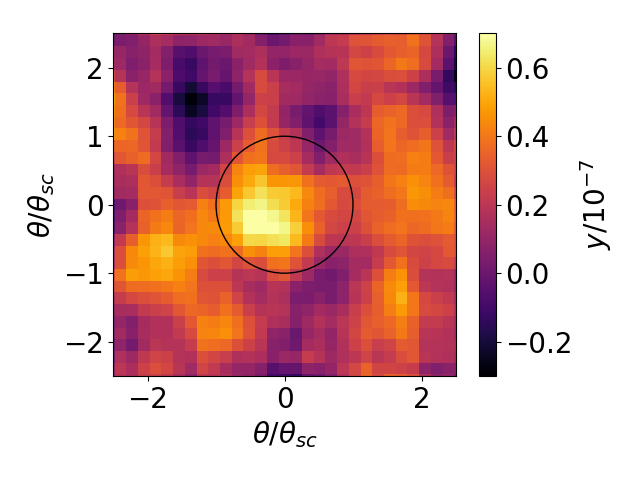}
    \includegraphics[width=0.33\linewidth]{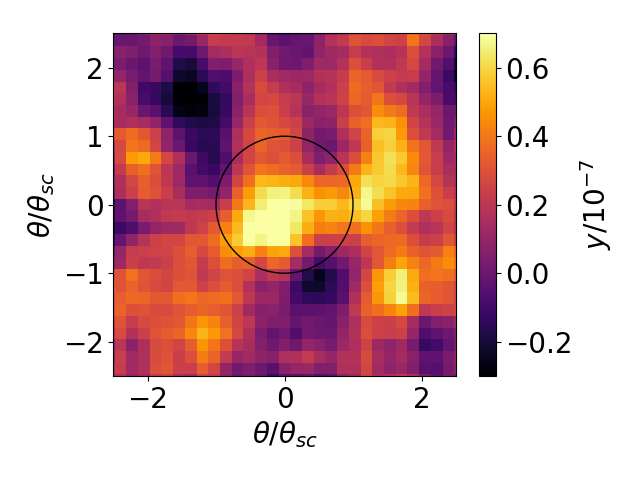}
    
    \includegraphics[width=0.33\linewidth]{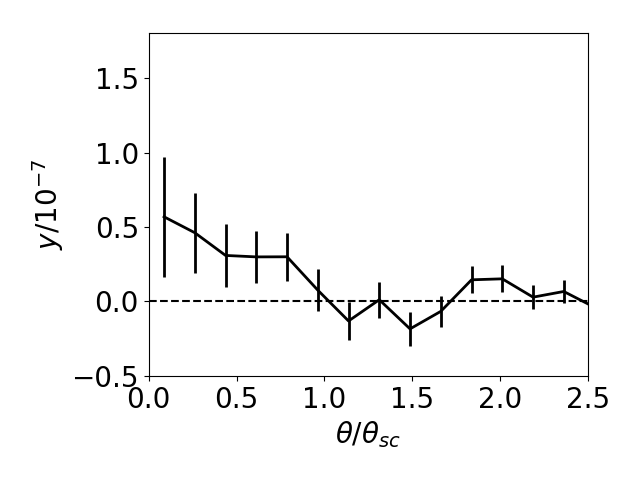} 
    \includegraphics[width=0.33\linewidth]{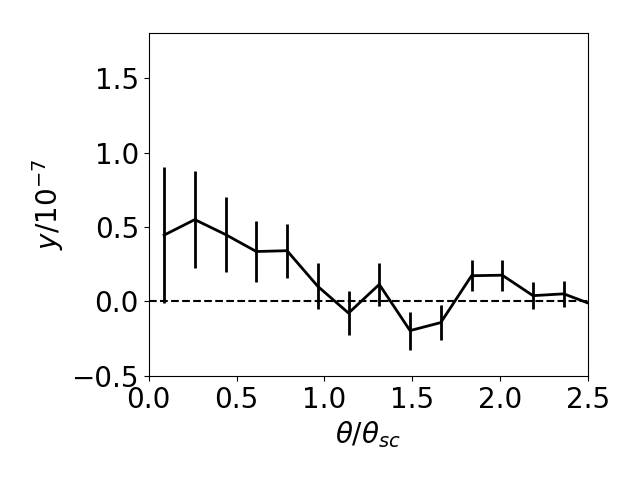}
    \includegraphics[width=0.33\linewidth]{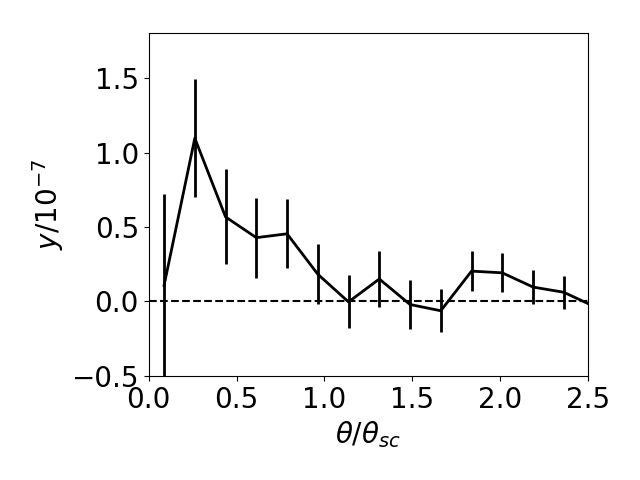}
    \caption{Stacked $y$ maps and corresponding $y$ profiles using different sizes of masks for galaxy clusters: $2 \times R_{500}$ ({\it left}), $3 \times R_{500}$({\it middle}) and $4 \times R_{500}$ ({\it right}). The mean $y$ values in the superclusters are estimated to be $y = (3.4 \pm 1.3) \times 10^{-8} \, (2 \times R_{500})$, $y = (3.5 \pm 1.4) \times 10^{-8} \, (3\times R_{500}),$ and $y = (3.1 \pm 2.0) \times 10^{-8} \, (4 \times R_{500}),$ respectively. In our analysis, we adopt a size of $3 \times R_{500}$ for the mask. }
    \label{f6}
    \end{figure*}

To check the validity of our mask, we change the mask size. In Fig.~\ref{f6}, the stacked $y$ maps and corresponding $y$ profiles using three different mask sizes are compared. We mask clusters using radii of $2 \times R_{500}$ ({\it left}), $3 \times R_{500}$ ({\it middle}), and $4 \times R_{500}$ ({\it right}). The bright central peaks associated with galaxy clusters disappear in all the cases, showing that these masks work well to remove the tSZ signal from galaxy clusters. However, a slight difference is seen. A residual around the center may remain in the left panel. While a slight excess around the center would be expected since this is 2D projection of 3D structure with an over-dense region around the core, it is better suppressed in the middle and right panels. On the other hand, bright spots re-appear in the right panel even with the larger mask. It shows that noises start to dominate due to overly the large mask (i.e., less regions available for stacking). Therefore, we adopt the size of $3 \times R_{500}$ for the mask of galaxy clusters. We discuss this more in Sect. \ref{subsec:uncertainties}. With this mask, the average tSZ signal ``outside of the mask'' in the superclusters is found to be $y = 3.5 \times 10^{-8}$.

\subsection{Signal-to-noise ratio}
\label{subsec:sn}
We assess the signal-to-noise ratio ($S/N$) of the tSZ signal from the superclusters. The $S/N$  for one supercluster can be estimated as a ratio of mean $y$ amplitude in a supercluster to $rms$ fluctuation of its background. The average $S/N$  of 669 superclusters is $\sim 0.8$. In the same manner, we assess the $S/N$  of our measured tSZ signal for the stacked superclusters in Sect. \ref{subsec:masksize} and find it to be $S/N = 2.3$.

\subsection{Null tests and error estimates}
\label{subsec:uncertainties}
\begin{figure} 
        \centering
        \begin{minipage}{1.0\linewidth}
        \includegraphics[width=\linewidth]{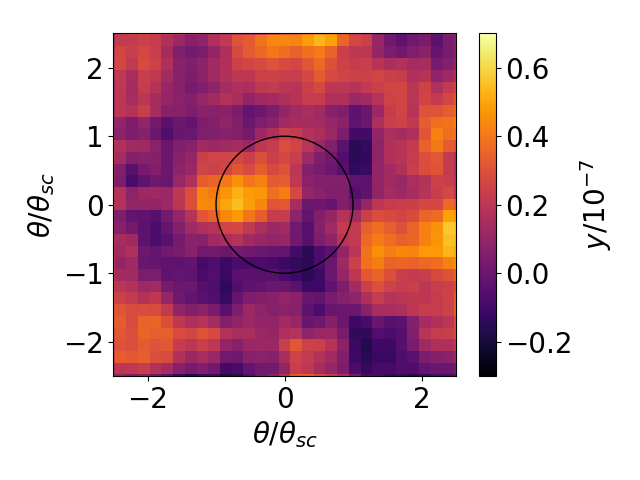}
        \end{minipage}
        \begin{flushleft}
        \begin{minipage}{0.8\linewidth}
        \includegraphics[width=\linewidth]{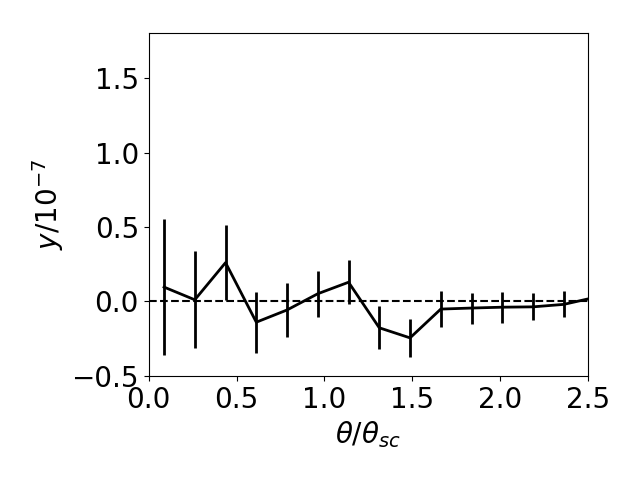}
        \end{minipage}
        \hspace{0.2\linewidth}
        \end{flushleft}
        \caption{{\it Upper}: Sample null map obtained by stacking the $y$ map against 669 superclusters randomly located on the sky.  {\it Lower}: Corresponding radial $y$ profile. }
        \label{f7}
    \end{figure}

\begin{figure}
    \centering
    \includegraphics[width=\linewidth]{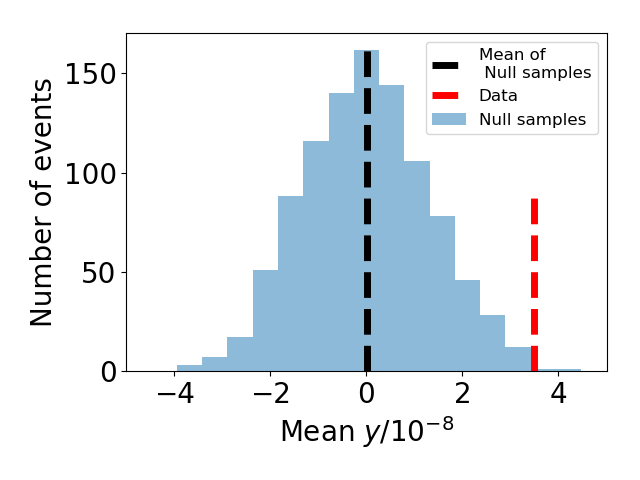}
    \caption{Distribution of mean $y$ amplitudes measured in 1000 null maps. The average of the distribution is shown with a {\it black dashed line} and the average $y$ signal from the data is shown with a {\it red dashed line}.}
    \label{f8}
    \end{figure}

To assess the significance of the tSZ signal and estimate its uncertainty, we perform a Monte Carlo-based null test. We move the center of each supercluster by a random angle in galactic longitude (while keeping the galactic latitude fixed, to avoid any systematic galactic background signal). For example, the center of one supercluster is changed from [galactic longitude, galactic latitude] = $[10^{\circ}, 60^{\circ}]$ to $[150^{\circ}, 60^{\circ}]$. We then stack the $y$ map at new ``random'' positions. We repeat the stacking process of our full samples 1000 times to determine the $rms$ fluctuations in the background (and foreground) sky.  Figure~\ref{f7} shows one of the 1000 stacked $y$ maps: the map has no discernible structure. We can use this ensemble of maps to estimate the uncertainty of the tSZ signal quoted above.  We find that the ensemble of the maps has a mean and standard deviation of $y = (0.0 \pm 1.3) \times 10^{-8}$ in Fig.~\ref{f8}.  Since the average signal in this null-test set of maps is consistent with zero, we conclude that our estimator is unbiased. Our measurement results in $y = (3.5 \pm 1.3) \times 10^{-8}$ at a significance of 2.7$\sigma$.

We also assess the significance of our measurement by a bootstrap resampling. For this, we draw a random sampling of 669 superclusters (790 before masking galaxy clusters) with replacement and re-calculate the average $y$ value for the new set of 669 superclusters. We repeat this process 1000 times and the bootstrapped data produce 1000 average $y$ values. Their average and $rms$ fluctuation are $y = (3.5 \pm 1.4) \times 10^{-8}$ at a significance of 2.5$\sigma$, which is consistent with the error estimate from the null test.

In order to check the independence of our results with respect to sizes of masks, we estimate the means and uncertainties of the tSZ signal for different masks in Fig.~\ref{f6}. These are estimated to be $y = (3.4 \pm 1.3) \times 10^{-8}$ for the mask of $2 \times R_{500}$, $y = (3.5 \pm 1.3) \times 10^{-8}$ for the mask of $3 \times R_{500}$, and $y = (3.1 \pm 2.0) \times 10^{-8}$ for the mask of $4 \times R_{500}$, respectively. These results are consistent, but the tSZ signal with a larger mask than $4 \times R_{500}$ is dominated by noise as described in Sect. \ref{subsec:masksize}.

\subsection{\bf Null hypothesis test}
\label{subsec:likelihood}
\begin{figure}
    \centering
    \includegraphics[width=\linewidth]{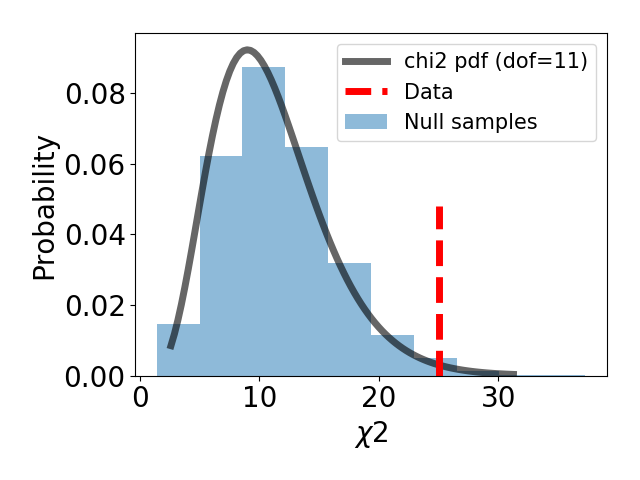}
    \caption{The $\chi^2$ distribution of 1000 null samples to null hypothesis ({\it blue}) is compared with a probability distribution with 11 degrees of freedom ({\it black}). The $\chi^2$ value of the data $y$ profile to the null hypothesis is shown as a {\it red dashed line}.}
    \label{f9}
    \end{figure}

From the null-test set of maps described above, we can generate an ensemble of 1000 null $y$ profiles and construct a covariance matrix to estimate the uncertainty of the data $y$ profile. We can assess a likelihood of the data $y$ profile to a null hypothesis with a chi-square test by computing
\beq
     \qquad \chi^2 = \sum_{i,j} (y(\theta_{i})-y_{mod}(\theta_{i}))^{T} (C^{-1})_{ij} (y(\theta_{j})-y_{mod}(\theta_{j})), 
\eeq
where $y(\theta_{i})$ is the $y$ value at $i$-th angular bin from the data and $y_{mod}(\theta_{i})$ is the corresponding value for a model ($y_{mod}(\theta_{i})=0$ for a null hypothesis). We verity that the $\chi^2$ distribution for the 1000 null-test sample is well described by a chi-squared distribution with 11 degrees of freedom (11 data points up to the scaled radius of $\sim$ 1.8) in Fig.~\ref{f9}. The $\chi^2$ value of the data $y$ profile to null hypothesis is estimated to be 25.0 for $dof=11$, providing a probability-to-exceed (PTE) of 0.009 which translates into a significance of 2.6$\sigma$.
    
We calculate the covariance matrix of the data $y$ profile in each radial bin from the 1000 average $y$ profiles obtained from the bootstrap resampling (see Sect. \ref{subsec:uncertainties}). Using this covariance matrix, we calculate a likelihood of the data $y$ profile to null hypothesis to be PTE = 0.014 which translates into a significance of 2.5$\sigma$. We adopt this covariance matrix to estimate the final uncertainty of the data $y$ profile due to instrumental noise and sky noise (i.e., cosmic variance and background subtraction errors). 

%________________________________________________________

\section{Interpretation}
\label{sec:interp}
%________________________________________________________

\subsection{Systematic errors}
\label{subsec:systematics}

First, we explore potential systematic effects in our measurements due to the \planck\ beam. Indeed the \planck\ beam may dilute the amplitude of our tSZ signal in superclusters in Fig.~\ref{f6}. However, the mean angular size of the superclusters is $\sim$ 2.8 deg and should have a minor effect compared to the \planck\ beam in the $y$ map of 10 arcmin.  Second, our measurements may be due to a beam-convolution of tSZ signal from galaxy clusters inside superclusters. While this ``leakage'' must be present at some level, if it were a dominant explanation for the residual signal, we would expect a significant difference for different sizes of masks. However, the fact that we see no significant dependence on the size of the mask suggests that the contribution from the leakage is not significant. Third, a contamination from cosmic infrared background
(CIB) in the \planck\ $y$ map is expected (as shown in \cite{planck2016-xxiii}) and it may mimic the measured signal. However, Fig.~ 14 in \cite{planck2016-xxiii} indicates that the CIB contamination in the \planck\ $y$ map at supercluster scales is small; it amounts to less than 10 \% in the power spectrum at the mean angular size of superclusters of order 2.8 deg. In addition, we explored the presence of the systematic effect in the reconstructed tSZ map using two \planck\ $y$ maps, MILCA and NILC, in Sect. \ref{subsec:gasprop}.   Finally, the physical interpretation of our measurements depends on supercluster morphologies as well as the distribution of intercluster gas in superclusters; neither is well constrained. The morphologies of the SDSS DR7 superclusters were studied in \cite{einasto2011} using a small selected sample of $\sim$ 35 large superclusters containing at least 300 member galaxies. The authors showed 2D and 3D distributions of galaxies and rich groups in the superclusters and found that most of them have filament-like overall shapes (see also Sect. \ref{subsec:gasprop}). 

\subsection{X-ray signal}
The thermal SZ effect has a linear dependence on gas density; on the other hand, the X-ray emission has a quadratic dependence. The X-ray emission is therefore important to break the degeneracy between density and temperature. Therefore, we stack the \rosat\ X-ray count rate maps \footnote{http://www.dlr.de/dlr/en/desktopdefault.aspx/tabid-10424/} around superclusters instead of the \planck\ $y$ map. We use the \rosat\ maps from the energy band (0.1-2.4 keV), hard energy band (0.5-2.4 keV), and soft energy band (0.1-0.4 keV), respectively. The galaxy clusters listed in Sect. \ref{subsec:groups} are all masked. We find the average X-ray signal to be consistent with zero. This result is to
be expected for such a low-density gas. While the degeneracy between density and temperature of the intercluster gas still remains, this result suggests that no significant signal is detected from X-ray-emitting systems such as galaxy clusters and that they are well masked.

\subsection{Gas properties}
\label{subsec:gasprop}

The Compton $y$ parameter in the direction $\hat{n}$, $y(\hat{n})$, is proportional to the integral of electron pressure $P_{\rm e}$ along the line of sight, 
\beq
        \qquad y(\hat{n}) = \frac{\sigma_{\rm T} }{m_{\rm e} {\it c}^2}
        \int P_{\rm e} \, (= n_{\rm e} k_{\rm B} T_{\rm e}) \, dl, 
        \label{eq:y}
\eeq
where $\sigma_{\rm T}$ is the Thomson cross section, $m_{\rm e}$ is the mass of electron, $c$ is the speed of light, $n_{\rm e}$ is the electron number density, $k_{\rm B}$ is the Boltzmann constant, and $T_{\rm e}$ is the electron temperature.

In general, the electron density at position ${\bf x}$ is given by
\beq
    \qquad n_{\rm e}({\bf x},z) = \overline{n}_{\rm e}(z)(1+\delta({\bf x})),
\eeq
where $\delta({\bf x})$ is the density contrast, and $\overline{n}_{\rm e}(z)$ is the mean electron density in the universe at redshift $z$,
\beq
    \qquad \overline{n}_{\rm e}(z) = \frac{\rho_{\rm b} ({\it z})}{\mu_{\rm e} m_{\rm p}}, 
\eeq
where $\rho_{\rm b} (z) = \rho_{\rm c} \Omega_{\rm b} (1+z)^3$ is the baryon density at redshift $z$, $\rho_{\rm c}$ is the present value of critical density in the universe, $\Omega_{\rm b}$ is the baryon density in unit of the critical density, $\mu_{\rm e} = \frac{2}{1+\chi} \simeq 1.14$ is the mean molecular weight per free electron for a cosmic hydrogen abundance of $\chi = 0.76$,  and $m_{\rm p}$ is the mass of the proton.

We can estimate physical properties of intercluster gas by considering a simple flat isothermal density distribution of gas (electrons) in superclusters with spherical shapes. Under these assumptions, the radial profile of the Compton $y$ parameter can be expressed as a geometrical projection of a 3D density profile with $n_{\rm e}(r,z)$
\beq
    \qquad y(r_{\bot}) = \frac{\sigma_{\rm T} k_{\rm B}}{m_{\rm e} c^2}  \int^{R}_{r_{\bot}} \frac{2r \, n_{\rm e}(r,z) \, T_{\rm e}(z)}{\sqrt{r^2 - r_{\bot}^2}} \, dr, 
    \label{eq:y-tangential}
\eeq
where $r_{\bot}$ is the tangential distance from a supercluster on the map and $R$ is the radius of a supercluster. Assuming a negligible evolution of intercluster gas (constant over-density $\delta_{\rm e}$ and constant temperature $T_{\rm e}$), 
\beqa
    \qquad  n_{\rm e}(r,z) &=& \frac{n_{\rm e}(r,z)}{\overline{n}_{\rm e}(z)} \, \overline{n}_{\rm e}(z) = \delta_{\rm e} \, \overline{n}_{\rm e}(z=0) \, (1+z)^3, \\
    T_{\rm e}(z) &=& T_{\rm e}.
\eeqa

However, our measured tSZ signal is not associated with the entire region of the superclusters due to the cluster mask used in our analysis. In addition, as introduced in Sect. \ref{subsec:systematics}, morphologies of superclusters are complicated and distribution of intercluster gas inside is not well constrained. We therefore make a correction to our model: we scale the amplitude of the $y$ profile for the superclusters using the supercluster volume used in our analysis.

The supercluster volume used in our analysis can be estimated from the volume given in the supercluster catalog. We remove the masked regions (cylindrical shape). However, since we do not have the information on the shapes of the superclusters, we instead calculate the unmasked volume for ``spherical supercluster''. Now the volume of the superclusters associated with the intercluster gas can be expressed by
\beq
    \qquad V_{\rm sc}^{\rm gas}(i) = f_{\rm sc}^{\rm um}(i) \times f_{\rm sc}^{\rm scat}(i) \times V_{\rm sc}^{\rm sph}(i), 
    \label{eq:Veff-sphere}
\eeq
where $f_{\rm sc}^{\rm um}(i)$ is the fraction of unmasked volumes of $i$-th supercluster relative to the spherical volume, $f_{\rm sc}^{\rm scat}(i)$ is the volume of $i$-th supercluster given by the catalog relative to the spherical volume and $V_{\rm sc}^{\rm sph}(i)$ is the spherical volume of $i$-th supercluster estimated using the radius, $r_{\rm sc}(i)$, defined by the maximum distance between its member galaxies ($V_{\rm sc}^{\rm sph}(i) = 4 \pi r_{\rm sc}(i)^3/3$). We use $f_{\rm sc}^{\rm um}(i) \times f_{\rm sc}^{\rm scat}(i)$ to correct the amplitude of the model $y$ profile of $i$-th supercluster.

\begin{figure}
    \centering
    \includegraphics[width=\linewidth]{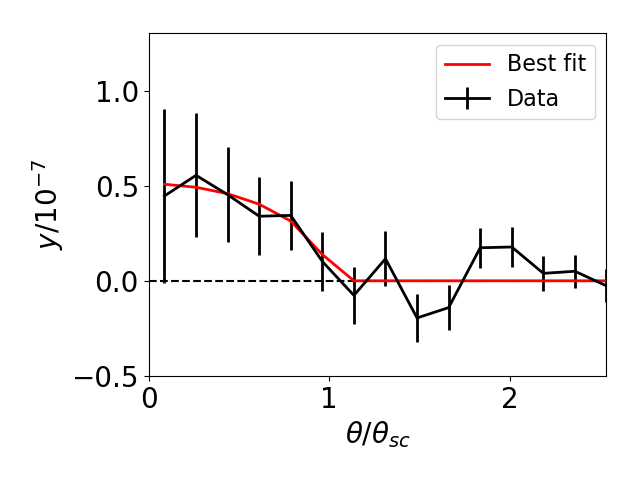}
    \caption{The average radial $y$ profile stacked at the positions of 669 superclusters using the \planck\ MILCA $y$ map ({\it black}), where three times the radius ($3 \times R_{500}$) of galaxy clusters is masked. The measured $y$ profile is fitted using a model with a constant over-density and constant temperature of gas in superclusters, described in Sect. \ref{subsec:gasprop} ({\it red}).}
    \label{f10}
    \end{figure}

For this model, we fit the data $y$ profile up to a scaled radius of 2.5 using the covariance matrix from the bootstrap resampling. The best-fit line is shown in Fig.~\ref{f10}.  Assuming a gas temperature of $T_{\rm e} = 8 \times 10^{6}$ K, which is the gas temperature in filaments between LRGs estimated from simulations in \cite{Tanimura2019}. This leads to 
\beq
    \qquad \delta_{\rm e}  \times \left(\frac{T_{\rm e}}{8 \times 10^{6} \; \rm{K}} \right) = 10.6 \pm 4.0, 
    \label{eq:overdensity-milca-sphere}
\eeq
with $\chi^2/dof=1.2$ for $dof$=14 (15 data point up to a scaled radius of 2.5 with one fit parameter for an amplitude of $y$ profile). 

In order to investigate a potential systematic error due to the \planck\ $y$ map, we repeat the analysis using the \planck\ NILC $y$ map. The $y$ profile obtained using the NILC $y$ map agrees with the profile resulting from the MILCA $y$ map. The derived value is $9.7 \pm 4.0$. We include this difference as a systematic error, resulting in 
\beq
    \qquad \delta_{\rm e}  \times \left(\frac{T_{\rm e}}{8 \times 10^{6} \; \rm{K}} \right) = 10.6 \pm 0.9 \, \pm 4.0 
    \label{eq:overdensity-milca-nilc-sphere}
,\eeq
which corresponds to 
\beq
    \qquad {\langle n_{\rm e} \rangle } \times \left(\frac{T_{\rm e}}{8 \times 10^{6} \; \rm{K}} \right) = (23.2 \pm 2.1 \, \pm 8.6 ) \, \times 10^{-7} [\rm{/cm}^3]. 
    \label{eq:numdensity-sphere}
\eeq

So far, we have assumed a uniform distribution of the gas in the superclusters. Now we estimate the gas density assuming that the measured tSZ signal is associated with filamentary structures inside superclusters. For this, we need to know the volume filling factor of these filamentary structures. However, the volume filling factor cannot be estimated directly from the data. We therefore use a cosmic value from Table 2 in \cite{Libeskind2018}, showing the volume fraction of cosmic-web structures such as knots, filaments, sheets, and voids from 12 different methods applied to an N-body simulation. We estimate the mean volume fraction of the internal filamentary structure of superclusters ($f_{\rm sc}^{\rm
fil}$) by summing the volume of the filaments and walls provided from six methods listed in \cite{Libeskind2018}. This is estimated to be $f_{\rm sc}^{\rm fil} \simeq 0.20$.

With this correction, the volume of the superclusters associated with the intercluster gas can be expressed by
\beq
    \qquad V_{\rm sc}^{\rm gas}(i) = f_{\rm sc}^{\rm fil} \times f_{\rm sc}^{\rm um}(i) \times f_{\rm sc}^{\rm scat}(i) \times V_{\rm sc}^{\rm sph}(i), 
    \label{eq:Veff-filament}
\eeq
and the gas properties can be re-estimated to be 
\beq
    \qquad \delta_{\rm e}  \times \left(\frac{T_{\rm e}}{8 \times 10^{6} \; \rm{K}} \right) = 53 \pm 5 \pm 20. 
    \label{eq:overdensity-milca-nilc-filament}
\eeq
This corresponds to 
\beq
    \qquad {\langle n_{\rm e} \rangle } \times \left(\frac{T_{\rm e}}{8 \times 10^{6} \; \rm{K}} \right) = (116 \pm 11 \, \pm 43 ) \, \times 10^{-7} [\rm{/cm}^3]. 
    \label{eq:numdensity-filament}
\eeq

\subsection{Baryon budget of intercluster gas in superclusters}
Diffuse gas in superclusters contributes to a total budget of baryons embedded in the large-scale structure. We estimate the contribution of our measurements to the total baryons using the model described in Sect. \ref{subsec:gasprop}. The total mass of the intercluster gas in the 669 superclusters can be evaluated using the over-density of electrons derived in Eq.\,\ref{eq:overdensity-milca-nilc-sphere} or Eq.\,\ref{eq:overdensity-milca-nilc-filament} and gas volumes in Eq.\,\ref{eq:Veff-sphere} or Eq.\,\ref{eq:Veff-filament}, 
\beq
    \qquad M_{\rm gas} = \sum^{N=669}_i\mu_{\rm e} \, \delta_{\rm e} \, \overline{n}_{\rm e}(z) \, V_{sc}^{gas}(i).
\eeq
We note that we obtain the same result in either case since the volume filling factor of filaments inside the superclusters cancels out.
The gas mass density can be calculated by setting
\beq
    \qquad \rho_{\rm gas} = \frac{M_{\rm gas}}{V_{\rm c}(z=0.2) \times f_{\rm SDSS}},  
\eeq
where $V_{\rm c}(z=0.2)$, which is the comoving volume at $z = 0.2$ (the maximum redshift in the supercluster samples) and $f_{\rm SDSS} \simeq 0.18, $ which is the fractional SDSS-DR7 survey field on the sky. The gas mass density relative to the critical density of the universe, $\Omega_{\rm gas}$, can be described as
\beq
    \qquad  \left(\frac{\Omega_{\rm gas}}{\Omega_{\rm b}} \right) \times \left(\frac{T_{\rm e}}{8 \times 10^{6} \; \rm{K}} \right) = 0.067 \pm 0.006 \, \pm 0.025.
\eeq
Assuming a gas temperature of $T_{\rm e} = 8 \times 10^{6} \, {\rm K}$, the total masses of the intercluster gas in the superclusters can be estimated to be $10^{12.8} - 10^{15.5} \, \msun$ with its average of $10^{14.5} \, \msun$. We note that the derived gas mass density is inversely proportional to the temperature of the gas. For example, the WHIM temperature in the northeast filament of A2744 is estimated to be $0.27^{+0.09}_{-0.05}$ keV ($\sim 3.1 \times 10^{6}$ K) by the \suzaku\ observations in \cite{hattori2017} using a two-component model of ICM and WHIM emissions. The temperature in filaments between CMASS galaxies with distances between 6 and 14 \mpc\ is also estimated to be $(2.7 \pm 1.7) \times 10^{6}$ [K] \citep{deGraaff2017}. With these lower temperatures, the gas densities can be estimated to be higher, that is, $\Omega_{\rm gas} / \Omega_{\rm b} \simeq$ 0.17 and 0.21, respectively. This may account for up to 44 - 52 \% of the missing baryons at low redshifts.

%________________________________________________________
\section{Discussion and conclusion}
\label{sec:discussion}
%________________________________________________________

In this paper, we present the first stacking analysis of the \planck\ tSZ maps around the superclusters identified by \cite{Liivamagi2012} from the SDSS DR7 galaxies. We detect the total tSZ signal from 790 superclusters at a significance of 6.4$\sigma$. This tSZ signal includes both intracluster and intercluster gas. To unveil a signal from diffuse low-density intercluster gas, we mask the galaxy groups/clusters identified by tSZ, X-rays, and optical surveys. We report the first detection of the tSZ signal from intercluster gas in superclusters with $y = (3.5 \pm 1.4) \times 10^{-8}$ at a significance of 2.5$\sigma$. The significance is estimated from null tests and bootstrap resampling and includes a possible systematic effect in the \planck\ tSZ maps.

In our analysis, we consider background-subtracted values of the tSZ signal. Strictly speaking, the measured tSZ signal in the superclusters should be associated to residual signal both in diffuse form and in low-mass systems residing in the superclusters. Estimating the relative contribution of both is complicated since it relates to the number of low-mass systems, their amount of hot gas, and the hypothesis on their clustering within superclusters. On one hand, if the low-mass systems are uniformly distributed, the background subtraction performed during our analysis should remove all the tSZ signal from low-mass systems, regardless of their hot gas content. On the other hand, the estimated level of tSZ signal in superclusters from clustered low-mass systems can be derived from the tSZ power spectrum. For all halos with masses below $10^{13.7}$ Msun, we compute the one- and two-halo terms of the tSZ power spectrum at the average angular scale of the  superclusters: $\sim$ 2.8 deg. We assume a Tinker et al (2008) mass function and a gas mass fraction from McGaugh et al (2010) for low-mass systems, and we find the ratio of two-halo to one-halo terms to $\sim 24$ \%,  corresponding to $\sim 25$\% of the measured tSZ signal. \footnote{The estimated value of 25\% would represent the maximum fraction of tSZ signal in superclusters associated with low-mass systems. The actual contribution is lower given that the measured tSZ signal is a background-subtracted value. Assuming the same gas mass fraction in low-mass systems as that in rich clusters increases the relative contribution from low-mass systems up to 35\% of the measured tSZ signal.}

Assuming a simple isothermal and flat density distribution of intercluster gas in superclusters, we estimate the product of over-density and temperature to be $\delta_e \times (T_{\rm e}/8 \times 10^{6} \, {\rm K}) =  10.6 \pm 0.9 \, (sys) \pm 4.0 \, (stat)$. The systematic error is estimated by comparing the analysis using two \planck\ $y$ maps, MILCA and NILC. This is re-evalulated to be $\delta_{\rm e} \times (T_{\rm e}/8 \times 10^{6} \, {\rm K}) =  53 \pm 5 \, (sys) \pm 20 \, (stat)$ with the assumption that our measured tSZ signal is associated with filamentary structures in superclusters. The degeneracy between the density and temperature can be broken using X-ray data. However, the X-ray signal around the superclusters from the \rosat\ X-ray maps is consistent with zero due to the low sensitivity of X-ray emission in low-density regions.  

Assuming a gas temperature of $T_{\rm e} = 8 \times 10^{6} \, {\rm K}$, estimated by simulations for filaments in \cite{Tanimura2019}, we find that the total gas mass density associated with our measurements corresponds to $\Omega_{\rm gas} / \Omega_{\rm b} \simeq 0.067 $. This accounts for $\sim$ 17 \% of missing baryons at low redhifts. The WHIM temperature in the northeast filament of A2744 is estimated to be $0.27^{+0.09}_{-0.05}$ keV ($\sim 3.1 \times 10^{6}$ K) by \suzaku\ observations in \cite{hattori2017} using a two-component model of ICM and WHIM emissions. The temperature in filaments between CMASS galaxies with distances between 6 and 14 \mpc\ is also estimated to be $(2.7 \pm 1.7) \times 10^{6}$ K \citep{deGraaff2017}. With these lower temperatures, the gas densities can be estimated to be higher, $\Omega_{\rm gas} / \Omega_{\rm b} \simeq$ 0.17 and 0.21, respectively. In conclusion, the derived gas density is inversely proportional to the gas temperature implying that our measurements of the intercluster gas may account for 17 - 52 \% of missing baryons at low redshifts.

Observations have been reported suggesting that large amounts of gas may be encompassed in a crowded environment of superclusters such as PLCK G214.6+37.0, Abell 2199 supercluster, and Corona Borealis supercluster. Along with these reports, our first statistical analysis of the tSZ signal from 689 superclusters allows us to explore the gas pressure. This allows us to derive, for the first time, a potential contribution of diffuse gas in superclusters to the total baryon budget. To probe such low-density regions, a better sensitivity to the tSZ signal would be needed in addition to X-ray data. Moreover, a better understanding of physical states of the gas in superclusters, especially its temperature, would be required to further identify the diffuse baryons and their contribution to the total baryon budget. It will be addressed using hydrodynamic simulations on very large scales. 

The future LSST \citep{lsst2008, lsst2009} and Euclid \citep{euclid2011} data will play an important role in more precise identification of galaxy groups/clusters as well as superclusters. Combination with other surveys, for example, eROSITA X-ray survey \citep{erosita2012} and Sunyaev-Zel’dovich surveys like ACTPpol \citep{niemack2010}, AdvACT \citep{advact2016}, or SPT-3G \citep{spt-3g2014} will help to unveil larger quantities of low-density gas in superclusters and to probe its physical properties. 

%________________________________________________________

\begin{acknowledgements}
This research has been supported by the funding for the ByoPiC project from the European Research Council (ERC) under the European Union's Horizon 2020 research and innovation programme grant agreement ERC-2015-AdG 695561. The authors acknowledge fruitful discussions with the members of the ByoPiC project (https://byopic.eu/team). We also thank P. Heinamaki and E. Saar for their suggestions. \\
This publication used observations obtained with {\planck} (\url{http://www.esa.int/Planck}), an ESA science mission with instruments and contributions directly funded by ESA Member States, NASA, and Canada. It made use of the SZ-Cluster Database (http://szcluster-db.ias.u-psud.fr/sitools/client-user/SZCLUSTER\_DATABASE/project-index.html) operated by the Integrated Data and Operation Centre (IDOC) at the Institut d'Astrophysique Spatiale (IAS) under contract with CNES and CNRS. This research has also made use of the VizieR database, operated at CDS, Strasbourg. The publication made use of the community-developed core Phyton package of astropy and numpy/scipy/matplotlib libraries. 
\end{acknowledgements}
%_____________________________________________________________________

\bibliographystyle{aa} % style aa.bst
\bibliography{bib} % your references Yourfile.bib

\end{document}